%
% the following is to use blackboard bold fonts --
\let\useblackboard=\iftrue
%
% activate this if you don't have them.
\let\useblackboard=\iffalse
%
% You might also need to remove this line.
%\newfam\black
%
\input harvmac.tex
%\input tables.tex
%
%%%%%%%%%%%%%%%%%%%%%%%%%%%%%%%%%%%%%%%%%%%%%%%%%%%%%%%%%%%%%%%
%The following lines are needed to insert the accompanying figures in
%the paper. If you do not have epsf, then comment out the line
% ``\input epsf'', and print the figures separately. The figures are at
%the end of the tex file, with instructions for their extraction.
\input epsf.tex
\ifx\epsfbox\UnDeFiNeD\message{(NO epsf.tex, FIGURES WILL BE
IGNORED)}
\def\figin#1{\vskip2in}% blank space instead
\else\message{(FIGURES WILL BE INCLUDED)}\def\figin#1{#1}\fi
\def\ifig#1#2#3{\xdef#1{fig.~\the\figno}
\midinsert{\centerline{\figin{#3}}%
\smallskip\centerline{\vbox{\baselineskip12pt
\advance\hsize by -1truein\noindent{\bf Fig.~\the\figno:} #2}}
\bigskip}\endinsert\global\advance\figno by1}
%%%%%%%%%%%%%%%%%%%%%%%%%%%%%%%%%%%%%%%%%%%%%%%%%%%%%%%%%%%%%%%%
\noblackbox
\baselineskip=12pt
\useblackboard
\message{If you do not have msbm (blackboard bold) fonts,}
\message{change the option at the top of the tex file.}

\font\blackboard=msbm10 scaled \magstep1
\font\blackboards=msbm7
\font\blackboardss=msbm5
%\newfam\black
\textfont\black=\blackboard
\scriptfont\black=\blackboards
\scriptscriptfont\black=\blackboardss

\else

\fi
% *************************************
%\draftmode
% *************************************

%
\def\yboxit#1#2{\vbox{\hrule height #1 \hbox{\vrule width #1
\vbox{#2}\vrule width #1 }\hrule height #1 }}
\def\fillbox#1{\hbox to #1{\vbox to #1{\vfil}\hfil}}
\def\ybox{{\lower 1.3pt \yboxit{0.4pt}{\fillbox{8pt}}\hskip-0.2pt}}

\def\comments#1{}

\def\Tr{{{\rm Tr\  }}}

\def\CN{{\cal N}}

\def\II{\relax{I\kern-.07em I}}

\def\ef{e\!f\!\!f}

\def\inbar{\,\vrule height1.5ex width.4pt depth0pt}
\def\IZ{\relax\ifmmode\mathchoice
{\hbox{\cmss Z\kern-.4em Z}}{\hbox{\cmss Z\kern-.4em Z}}
{\lower.9pt\hbox{\cmsss Z\kern-.4em Z}}
{\lower1.2pt\hbox{\cmsss Z\kern-.4em Z}}\else{\cmss Z\kern-.4em
Z}\fi}
\def\IB{\relax{\rm I\kern-.18em B}}
\def\IC{{\relax\hbox{$\inbar\kern-.3em{\rm C}$}}}
\def\ID{\relax{\rm I\kern-.18em D}}
\def\IE{\relax{\rm I\kern-.18em E}}
\def\IF{\relax{\rm I\kern-.18em F}}
\def\IG{\relax\hbox{$\inbar\kern-.3em{\rm G}$}}
\def\IGa{\relax\hbox{${\rm I}\kern-.18em\Gamma$}}
\def\IH{\relax{\rm I\kern-.18em H}}
\def\IK{\relax{\rm I\kern-.18em K}}
\def\IP{\relax{\rm I\kern-.18em P}}
\def\pp{{\relax{=\kern-.42em |\kern+.2em}}}
%\def\IX{\relax{\rm X\kern-.01em X}}
%this doesn't work

\font\cmss=cmss10 \font\cmsss=cmss10 at 7pt
\def\IR{\relax{\rm I\kern-.18em R}}

\def\Tr{{\rm Tr\ }}

%%

%
%Journal macros
%

%
%\baselineskip 22pt
%
\lref\yanki{A. A. Tseytlin and S. Yankielowicz, Nucl.Phys. B541 (1999) 145, hep-th/9809032.}
\lref\kiri{E. Kiritsis and T.R. Taylor, ``Thermodynamics of D-brane Probes'', hep-th/9906048;
E. Kiritsis, ``Supergravity, D-brane Probes and thermal super Yang-Mills: a comparison, 
hep-th/9906206.}
\lref\juanprobes{J. Maldacena, ``Probing Near Extremal Black Holes with D-branes'', 
hep-th/9705053; ``Branes Probing Black Holes'', Nucl.Phys.Proc.Suppl. 68 (1998) 17,
hep-th/9709099.}
\lref\jw{J. Maldacena, ``Wilson loops in large N field theories'', 
Phys.Rev.Lett. 80 (1998) 4859, hep-th/9803002.}
\lref\reyw{S.-J. Rey and J. Yee, ``Macroscopic Strings as Heavy Quarks 
of Large N Gauge Theory and Anti-de Sitter Supergravity'', hep-th/9803001.}
\lref\ya{A. Brandhuber, N. Itzhaki, J. Sonnenschein and S. Yankielowicz,
``Wilson Loops in the Large N Limit at Finite Temperature'', Phys.Lett. 
B434 (1998) 36, hep-th/9803137.}
\lref\wfirst{E. Witten, ``Anti De Sitter Space And Holography'', 
Adv.Theor.Math.Phys. 2 (1998) 253, hep-th/9802150.}
\lref\wsec{E. Witten, ``Anti-de Sitter Space, Thermal Phase Transition, 
And Confinement In Gauge Theories'', Adv.Theor.Math.Phys. 2 (1998) 505,
hep-th/9803131.}
\lref\hp{S. W. Hawking and D. Page, ``Thermodynamics of Black Holes
in Anti-de Sitter Space'', Comm.Math.Phys. 87 (1983) 577.}
\lref\bg{T. Banks and M. B. Green, ``Non-perturbative Effects in AdS5*S5 
String Theory and d=4 SUSY Yang-Mills'', JHEP 9805 (1998) 002, hep-th/9804170.}
\lref\theisen{S. Forste, D. Ghoshal and S. Theisen, ``Stringy Corrections 
to the Wilson Loop in N=4 Super Yang-Mills Theory'', hep-th/9903042.} 
\lref\theisenwlt{S.-J. Rey, S. Theisen and J.-T. Yee, ``Wilson-Polyakov 
Loop at Finite Temperature in Large N Gauge Theory and Anti-de Sitter
Supergravity'', Nucl.Phys. B527 (1998) 171, hep-th/9803135.}
\lref\ak{A. Brandhuber and K. Sfetsos, ``Wilson loops from multicentre and 
rotating branes, mass gaps and phase structure in gauge theories'', 
hep-th/9906201.}
\lref\elliptic{P. F. Byrd and M. D. Friedman, ``Handbook of Elliptic
Integrals for Engineers and Scientists'', second edition, 1971 
Springer-Verlag.}
\lref\rot{P. Kraus, F. Larsen and S. P. Trivedi, ``The Coulomb Branch 
of Gauge Theory from Rotating Branes'', JHEP 9903 (1999) 003, hep-th/9811120.}
\lref\rots{K. Sfetsos, ``Branes for Higgs phases and exact conformal 
field theories'', JHEP 9901 (1999) 015, hep-th/9811167.}
\lref\roberto{R. Emparan, ``AdS/CFT Duals of Topological Black Holes and 
the Entropy of Zero-Energy States'', JHEP 9906 (1999) 036, hep-th/9906040.}
\lref\rcm{R. Emparan, C. V. Johnson and R. C. Myers, ``Surface Terms as 
Counterterms in the AdS/CFT Correspondence'', hep-th/9903238.}
\lref\gary{G. T. Horowitz and R. C. Myers, ``The AdS/CFT Correspondence and a New Positive Energy Conjecture for General Relativity'',
Phys. Rev. D59 (1999) 2605, hep-th/9808079.}
\lref\juani{J. Maldacena, ``The large N limit of superconformal field theories
and supergravity'', Adv. Theor. Math. Phys. 2 (1998) 231, hep-th/9711200.}
\lref\garyi{ Gary T. Horowitz and Simon F. Ross, ``Possible Resolution of Black Hole Singularities from Large N Gauge Theory'', JHEP 9807 (1998) 014,
hep-th/9803085.}
\lref\review{O. Aharony, S. S. Gubser, J. Maldacena, H. Ooguri and Y. Oz,
``Large N Field Theories, String Theory and Gravity'', hep-th/9905111.}
\lref\kapusta{J. I. Kapusta, ``Finite Temperature Field Theory'', Cambridge
Monographs on Mathematical Physics (1989).}
\lref\sw{L. Susskind and E. Witten, ``The Holographic Bound in Anti-de Sitter Space'', hep-th/9805114.}
\lref\bcm{C. P. Burgess, N. R. Constable and R. C. Myers, ``The Free Energy of N=4 Super-Yang-Mills
Theory and the AdS/CFT Correspondence'', hep-th/9907188.}
\lref\mann{R. B. Mann, ``Topological Black Holes -- Outside Looking In'', gr-qc/9709039.}
\lref\freund{P.G.O. Freund and M. A. Rubin, ``Dynamics of Dimensional Reduction'',
Phys.Lett.97B: (1980) 233.}
\lref\pol{J. Polchinski, ``String Theory'', vol. 2, Cambridge University 
Press, 1998.}
\lref\re{R. Emparan, ``AdS Membranes Wrapped on Surfaces of Arbitrary Genus'', Phys.Lett. B432 
(1998) 74, hep-th/9804031.}
\lref\odintsov{S. Nojiri and S. D. Odintsov, ``Running Gauge Coupling and 
Quark-Antiquark Potential in Non-SUSY Gauge Theory at Finite
Temperature from IIB SG/CFT correspondence'', hep-th/9906216.}
\lref\bir{D. Birmingham, ``Topological Black Holes in Anti-de Sitter Space'',
Class.Quant.Grav. 16 (1999) 1197, hep-th/9808032.}
\lref\bd{N. D. Birrell and P. C. W. Davies, ``Quantum Fields in Curved Space'',
Cambridge University Press, 1982.}
\lref\duff{M. J. Duff, ``Twenty Years of the Weyl Anomaly'', Class. 
Quant. Grav. 11 (1994) 1387, hep-th/9308075 and references therein.}

%
%%%%%%%%%%%%%%%%%%%%%%%%%%%%%%%%%%%%%%%%%%%%%%%%%%%%%%%%%%%%%%%%%%%%%%
%
\Title{ \vbox{\baselineskip12pt\hbox{hep-th/9908010}
\hbox{TUW-99-17}
}}
{\vbox{
\centerline{Probing the Strong Coupling Limit of Large $N$} 
\vskip3mm
\centerline{SYM on Curved Backgrounds}}}

\centerline{Karl Landsteiner and Esperanza Lopez}
\medskip
\centerline{Institut f\"ur theoretische Physik}
\centerline{Technische Universit\"at Wien}
\centerline{Wiedner Hauptstra{\ss}e 8-10}
\centerline{A-1040 Wien, Austria}
\medskip
\centerline{\tt landstei@brane.itp.tuwien.ac.at}
\centerline{\tt elopez@tph44.tuwien.ac.at}

\vskip15mm

\centerline{\bf Abstract}

%\vskip5mm
\baselineskip=14pt
According to the AdS/CFT correspondence, the strong coupling limit of 
large $N$ ${\cal N}\!=\!4$ supersymmetric gauge 
theory at finite temperature is described by asymptotically anti de Sitter black holes. 
These black holes exist with planar, spherical and hyperbolic horizon 
geometries. We concentrate on the hyperbolic and spherical cases and 
probe the associated gauge theories with D3-branes and Wilson loops. 
The D3-brane probe reproduces the coupling of the scalars in the gauge 
theory to the background geometry and we find thermal stabilization 
in the hyperbolic case. We investigate the vacuum expectation
value of Wilson loops with particular emphasis on the screening length at 
finite temperature.  
We find that the 
thermal phase transition of the theory on the sphere is not related to 
screening phenomena.  
\vfill
\Date{\vbox{\hbox{\sl August, 1999}}}

%References

\newsec{Introduction}
Maldacena's conjecture \juani\ states that $SU(N)$ $\CN=4$ supersymmetric
gauge theory in four dimensions is dual to type IIB string theory on an 
$AdS_5\times S^5$ background. This conjecture is obtained by considering a 
collection of $N$ D3-branes in type IIB string theory. 
The effective theory describing the physics of the
D3-branes is $\CN=4$ $SU(N)$ Yang-Mills. D3-branes can also be described as 
classical black hole solutions in IIB supergravity. 
Taking $\alpha' \rightarrow 0$ and at the same time keeping the mass of the 
strings stretched between the D3-branes finite defines the near horizon limit 
of the supergravity solution. On the other hand one is keeping only the 
ground states of the open strings. This gives $\CN=4$ gauge theory as the 
exact theory (not only the effective theory for low energies) and the 
$AdS_5\times S^5$ background as the near horizon limit in the supergravity 
solution.

The same arguments apply not only to extremal but also to near-extremal D3-branes \garyi. 
One obtains then asymptotically anti de Sitter black holes with planar horizon geometry.
The euclidean continuation of these black hole solutions is conjectured to describe
a thermal state of the $\CN=4$ gauge theory with a temperature corresponding to the
Hawking temperature of the black hole \refs{\juani,\wsec}. Much work has been devoted to the study of this
duality, see e.g. \review\ for a review.

The supergravity 
solution is however only an approximation to string theory and corresponds to the
large $N$ and strong coupling limit in the dual gauge theory. A direct comparison of 
quantities calculated at weak coupling in
the gauge theory with the corresponding results in the supergravity theory is therefore
rather difficult. However many quantities calculated at strong coupling have been shown to be 
consistent at the
qualitative level with general expectations from gauge theories.
Among the quantities that have been studied for near-extremal D3-branes are
effective potentials obtained through D3-brane probes \refs{\yanki, \kiri} 
and vacuum expectation values of Wilson loops corresponding to the action of 
strings ending on the conformal boundary of the AdS black hole \refs{\jw, \reyw, \ya, \theisenwlt, \ak}.

Black holes in AdS space can have spherical, planar or hyperbolic horizon geometries. 
Although only the planar
black holes can be directly related to D3-branes in a thermal state, one
expects that also the spherical and hyperbolic black holes describe gauge 
theories. More specifically in \wsec\ it was conjectured that spherical black 
holes describe
$\CN=4$ gauge theories at finite temperature on $S^3$. One also expects that 
AdS black holes with hyperbolic horizon describe $\CN=4$ gauge theory on the
hyperbolic three-plane $H^3$. We will further investigate these conjectures.
In section 2 we will compute the static potential for a test D3-brane in the
background of spherical or hyperbolic AdS black holes. We find that at large
distances the static potential reproduces the tree level coupling of the 
scalars
to the background geometry in the gauge theory. Since in the hyperbolic case
this coupling induces a negative mass term one would expect the gauge theory 
to be unstable. The D3-brane potential at finite temperature grows however 
for small
distances up to a maximum value and only for large distances reproduces the 
negative mass term in the gauge theory. At zero temperature there is however
no such stabilization and one finds only the tree level negative mass term 
\foot{Similar behavior of the potential of M2-branes in the background of 
non-extremal D-branes with hyperbolic horizon has been found in \re.}. 

In section 3 we turn to the properties of Wilson loops. We find screening at 
finite temperature on the sphere and at zero temperature a potential that is 
very close
to the form of the potential at weak coupling. We also speculate about the
relation of the screening length at finite temperature and the thermal phase
transition between the spherical AdS black hole and empty AdS space.
In the hyperbolic case we find screening even at zero temperature.

\newsec{Probing AdS Black Holes with D-Branes}

We begin with the relevant bosonic part 
of the type IIB action\foot{We take a Yang-Mills term 
for the five-form action with the convention of imposing self-duality at the 
level of the equations of motion.}
\eqn\action{ S_{IIB} = 
{1\over 2 \kappa_{10}^2} \int d^{10}x \sqrt{-g} \left( R - {1\over 4.5!} 
F_5^2\right) \,. 
}
The ten-dimensional spacetime will be the product manifold of $S^5$ and a 
five-dimensional AdS black hole. The flux of the five-form field strength 
through $S^5$ is quantized in terms of the D3-brane charge as 
$\int F = 2 \kappa_{10}^2 \mu_3 N$, with 
$\kappa_{10}^2 = 64 \pi^7 g_s \alpha'^4$ and 
$\mu_3^2 = {\pi\over\kappa_{10}^2}$.
The metric of the euclidean AdS black holes is
\eqn\adsbh{ 
ds^2 = \ell^2 \left[ \left(u^2+k-{\mu \over u^2}\right) d\tau^2 + 
{du^2\over u^2+k-{\mu \over u^2}} + u^2 \, d\Sigma_{3,k}^2 \, \right]\,,
}
where $\ell$ is the radius of curvature of the asymptotic AdS geometry 
and $\mu$ is proportional to the black hole mass. The discrete parameter 
$k$ can take the values $(1,0,-1)$
corresponding to the spherical, planar or hyperbolic cases respectively. 
In the following we will always consider $k\!=\!1$ or $k\!=\! -1$. 
$d\Sigma_{3,k}^2$ denotes then the metric on the unit 3-sphere or 
3-hyperboloid 
\eqn\threem{ 
\eqalign{
d \Sigma_{3,1}^2 = & d \theta^2 + \sin^2 \! \theta \, d \Omega_2^2 ,\cr
d \Sigma_{3,-1}^2 = & d \theta^2 + \sinh^2 \! \theta \, d \Omega_2^2 .\cr}
}
The geometry defined in \adsbh\ is free of conical singularities if
the euclidean time coordinate $\tau$ is periodically identified, with 
period 
\eqn\defbeta{ 
\beta = { 2\pi u_+ \over 2u_+^2+k }\,,
}
where $u_+$ is the horizon radius 
\eqn\defrp{ 
u_+ = \sqrt{- {k \over 2} + \sqrt{{1 \over4} + \mu}}\,.
}
The temperature associated to \adsbh\ is $T={1\over \beta}$. 
Since the boundary of asymptotically AdS metrics is only defined 
up to conformal transformations \wfirst, we have chosen \adsbh\ such that
the boundary 3-sphere or 3-hyperboloid has radius one. Although an
arbitrary radius $R$ can be set by redefining $u \! = \! R {\tilde u}$ and 
$\tau \! =\! {\tilde \tau}/R$, the thermodynamics will only depend on the
ratio $\beta_{S^1}/R$ given by \defbeta.

The black hole metrics \adsbh\ with $k=1$ only exist for $T\geq T_{min} = 
{\sqrt{2}\over \pi}$, which corresponds to $\mu={3\over 4}$.
For each temperature bigger than $T_{min}$ there are two solutions for $\mu$, 
one bigger than 
$3\over 4$ and one smaller. We will only consider mass parameters 
$\mu \geq {3 \over 4}$.
Black holes with smaller values of $\mu$ have negative specific heat.
They decay due to Hawking radiation and do not correspond 
to a stable thermal state in the gauge theory. There is furthermore a 
phase transition at $\mu_{crit}=2$, or equivalently
\eqn\tcrit{
T_{crit}= { 3 \over 2 \pi }.
}
For temperatures below $T_{crit}$ the black hole solution is unstable 
due to tunneling to empty 
AdS-space \hp. The gauge theory interpretation is that of a 
confining/deconfining phase transition \refs{\wfirst, \wsec}. The free energy 
of the black hole solutions scales like $N^2$, consistent with the 
expected deconfinement at finite temperatures in the gauge theory. 
For temperatures $T< T_{crit}$ the field theory dual is empty AdS.
Its free energy scales like $N^0$, signalling confinement. We will discuss 
this phase transition in more detail in section 3. 

For $k=-1$ the horizon radius is defined even for negative mass 
parameters down to $\mu = -{1\over 4}$. Lorentzian black holes 
with negative mass parameter have also an inner 
horizon and a timelike singularity such that their Penrose diagram is that
of the Reissner-Nordstrom AdS solution \mann. At the extremal case 
$\mu=- {1\over 4}$ inner and outer horizon coincide. The nature of the 
horizon changes, it becomes a bifurcate horizon with no particular 
temperature associated to it. Thus the extremal solution is somewhat
similar to empty AdS in the Poincar{\'e} patch and can be associated
with the zero temperature ground state of ${\cal N}=4$ Yang-Mills
on the hyperboloid \refs{\bir,\rcm}. The fact that the groundstate is given 
by a 
solution with negative mass parameter and its implications concerning the 
entropy of the corresponding gauge theory have been discussed in \roberto\
and the field theory 1-loop calculation of the free energy has been performed 
recently in \refs{\roberto,\bcm}.

Taking the metrics \adsbh\ as input for a Freund-Rubin like
ansatz \freund\ we find that the length scale $\ell$ is determined by the five-form 
flux through $S^5$ as $\ell^4 = 4\pi g_s N \alpha'^2$ and the potential 
of the five-form field strength is given by
\eqn\potfour{ 
A_4 = -\ell^4 u^4  d\tau d\Sigma_{3,k} \,,
}
where $d \Sigma_{3,k}$ is the unit volume form on either $S^3$ or $H^3$.
Notice that the metrics \adsbh\ are not derived as near-horizon limits of 
D3-branes on an asymptotically flat space.

${\cal N}=4$ Yang-Mills theory in flat space has a Coulomb branch of vacua 
on which the gauge group is broken to a smaller group of equal rank. In
D-brane language this translates into forces between 
parallel extremal flat D3-branes being zero. Separating D3-branes in the
transverse space dimensions corresponds to turn on Higgs expectation
values in the gauge theory. At finite temperature the field
theory develops an effective potential which lifts the Coulomb branch
and drives the system back to symmetry 
restauration \kapusta. Accordingly, forces between near-extremal D3-branes
do not cancel. A symmetry breaking pattern $U(N+1)\rightarrow U(N)\times 
U(1)$ is given in the gravity description by separating a single D3-brane 
from the rest. The effective action of such a configuration can be 
computed as the action of the single D3-brane in the background 
of the other $N$ branes 
\eqn\bi{
S= T_3 \int \left( \sqrt{\hat g} + A_4 \right),
}
where $T_3$ is the D3-brane tension, $\hat g$ the induced metric on the 
probe world-volume
and $A_4$ is the background value of the type IIB four-form potential.
This calculation has been done in \yanki\ and \kiri\ and qualitative 
agreement with the expectations from gauge theory has been found. 
In this section we want to extend this analysis to the spherical and
hyperbolic cases. The main difference with the flat case is that 
the ${\cal N}=4$ scalars can get mass due to a non-minimal coupling
to the boundary geometry,  
\eqn\confscalar{ 
S_{SYM}={1\over g^2_{YM}}\int d^4x \sqrt{g} \;\, \Tr \!\! \left( 
{1\over 4} F_{\mu\nu} F^{\mu\nu}+{1\over 2} D_\mu \Phi D^\mu 
\Phi + {1\over 2} \xi R \Phi^2 +\cdots\right)\,.
}
Therefore we expect a non-zero effective potential even at zero 
temperature. Since the boundary geometry is only defined as a conformal 
class, the ${\cal N}=4$ scalars should be conformally coupled, i.e. 
$\xi=1/6$.

We consider a test D3-brane at a distance $u$ from the origin. 
Substituting \adsbh\ and \potfour\ in the
Born Infeld action \bi\ we obtain \foot{In \refs{\yanki, \kiri}
the authors worked with asymptotically flat geometries describing 
near-extremal 3-branes solutions. In defining the probe action, 
they subtracted the 3-brane volume at infinity. We are directly 
working in asymptotically AdS spaces and the D3-brane action 
at infinity does not become a constant therefore no such subtraction 
can be done. Our expression for $k=0$ coincides with \yanki\ 
and up to a constant also with \kiri.}
\eqn\actionDthree{ 
S = \beta T_3 V_3 \ell^4  u^4 \left( \sqrt{1+{k \over u^2} 
- { \mu \over u^4}} - 1\right) \, ,
}
with $V_3$ being the volume of the unit 3-sphere or 3-hyperboloid.
For distances $u >> {1\over T}$, the potential $V=T \, S$ derived from
\actionDthree\ becomes 
\eqn\largeu{ 
V ={k\over 4 \pi^2} N V_3 \, u^2 \,,
}
where we used $T_3\! =\! {1\over (2 \pi)^3 g_s}\!=\! 
{N\over 2 \pi^2 \ell^4}$ \pol. 
To compare this to the gauge theory in \confscalar\ we introduce the 't 
Hooft coupling $\lambda = 2 g^2_{YM} N$ and identify it on the
the supergravity side through $\ell^4 = \lambda \alpha'^2$.
The radial coordinate $u$ is then related to the Higgs expectation 
value by ${u \over 2 \pi}= {\Phi \over \sqrt{\lambda}}$ \refs{\juanprobes, 
\yanki, \kiri}. The curvature of a unit 3-sphere or 3-hyperboloid
is $R=6 k$. Substituting these values in \confscalar\ with $\xi=1/6$, 
we obtain precisely \largeu. Thus we recover at large $u$ precisely the 
tree-level coupling of the scalar fields to the background curvature in 
the gauge theory. At first glance it seems surprising that this term does 
not suffer some non-trivial renormalization since supersymmetry is broken 
by both the background geometry and the temperature. At large $u$
we are however probing the ultraviolet properties of the gauge theory.
Since ${\cal N}=4$ Yang-Mills on $S^1 \! \times \! S^3$ or $S^1 \!
\times \! H^3$ has vanishing conformal anomaly, in the large $u$ limit we 
expect to probe the bare couplings of the ${\cal N}=4$ gauge theory. This 
result is an example for the suggestion that large distances in the bulk 
correspond to small distance scales on the boundary \sw.

It is interesting to analyze the large $u$ behaviour of the potential
when the dual gauge theory lives on an space with non-vanishing
conformal anomaly. We will consider as an example $AdS_5$ with $S^4$
boundary of radius one. The metric is
\eqn\cometric{
ds^2 = \ell^2 \left(  
{du^2\over u^2+1} + u^2 \, d\, \Omega_4^2 \, \right)\,.
} 
Using this metric as input for a Freund-Rubin like
ansatz, we obtain its associated four-form potential 
\eqn\copot{
A_4 = -\ell^4  \left[\Bigl(u^3 - {3 \over 2}u \Bigr)\sqrt{1 + u^2} + 
    {3 \over 2}\,{\rm arcsh}\, u \right]   d {\rm Vol}_{S^4}.
}
Substituting these values in \bi\ and expanding for large $u$ we get
\eqn\colargeu{
V ={1 \over 2 \pi^2}N V_4  \left( u^2- {3 \over 2} \, {\rm log}\, u 
\right) \,,
} 
where $V_4$ is the volume of $S^4$. The curvature of a four-sphere of 
radius one is $12$. The first term in \colargeu\
reproduces the conformal coupling of the scalars to the background 
curvature with the same identification between $u$ and $\Phi$ as before. 
We associate the second term to the conformal anomaly. The trace
anomaly of ${\cal N}=4$ on $S^4$ is $T \!=\!
{3 N^2 \over 8 \pi^2}$ \bd\ and thus the effective action picks up
a logarithmic term \duff
\eqn\logdiv{
{3 \over 8 \pi^2}\, N^2  V_4 \, {\rm log} \, \epsilon \,,
}
with $\epsilon$ playing the part of the momentum cutoff. The coefficient that
multiplies the logarithm in \colargeu\ is precisely the leading $O(N)$ 
term in $T_{N}-T_{N+1}$. The variable $u$ sets the energy scale in the field
theory at which the gauge group is Higgsed from $SU(N\!+\!1)$ to $SU(N)$. It
is therefore $u$ what substitutes the cutoff $\epsilon$ in \colargeu.

The identification of $u$ with the vacuum expectation value of the scalar 
fields is only correct at large $u$. It has been argued \juanprobes\ that 
in order to compare radial distances in supergravity with energy scales in the
gauge theory one should use an isotropic coordinate defined by 
\eqn\isocooi{ 
\rho = \exp \int {du \over u^2 + k - {\mu \over u^2}} = 
\left( u^2+{k\over 2} + \sqrt{u^4 + k u^2 - \mu} \right)^{1\over 2},
}
or equivalently
\eqn\isocooii{ u^2 = { \rho^4 + \mu + 1/4 \over 2 \rho^2} - {k\over 2} \,.
}
In terms of the new coordinate $\rho$, the difference between the D3-brane 
potential at infinity and at the horizon is
\eqn\deltaV{V(\rho_+)-V(\infty) = {V_3 N \over 4 \pi^2} \left(- k 
{\rho^2 \over 2} \big|_{\rho\rightarrow \infty}+ \rho_+^4 - 2 k 
\rho_+^2\right) \,.
}
The action of the euclidean black hole solutions \adsbh\ has been 
calculated in \refs{\bir,\rcm}. Using their results and our coordinate, 
we find for the free energy
\eqn\freeE{ 
F_N = {V_3 N^2 \over 8 \pi^2} ( \rho_+^4 - 2 k \rho_+^2 ) \,.
}
If we add one D3-brane the free energy changes by an amount 
$F_{N+1} - F_N$. The leading $O(N)$ term of this change is precisely given 
by \deltaV\ after discarding the quadratically divergent term. The quadratic 
divergence does not influence the thermodynamics since it is temperature 
independent. For this reason we do not see this contribution in the change
of the free energy.  

It is convenient to introduce an ``effective temperature'' defined by
\eqn\efftempi{\rho_+ = \pi T_{\ef},} 
\eqn\efftempii{ 
T_{\ef} = {T\over \sqrt{2}} 
\left( 1 + \sqrt{1-{2 k \over \pi^2 T^2 } }\right)^{\! 1\over2} \,.
}
In analogy with \yanki\ and \kiri\ we then introduce the scalar mass parameter 
$M$ by
\eqn\defM{ M = \rho - \rho_+ \,.}
The static D3-brane potential takes now the form
\eqn\staticpot{ V(M) = {\pi^2 T_{\ef}^4 V_3 N \over 4} \left[ 1 - 
{1\over \bigl(1\!+\!{M\over \pi T_{\ef}}\bigr)^4} + 
{k\over 2 \pi^2 T_{\ef}^2} \left( 
{3\over \bigl( 1\!+\!{M\over \pi T_{\ef}}\bigr)^2} +  \Bigl(1\!+\!
{M\over \pi T_{\ef}}\Bigr)^2 - 4 \right) \right] \,,}
where we chose to subtract a constant such that it vanishes at the horizon.  
The first two terms in \staticpot\ reproduce the potential obtained for
planar horizon in \refs{\yanki, \kiri} by substituting $T_{\ef}$ with $T$ 
\foot{Note however that the potentials in \refs{\yanki, \kiri} have not been 
chosen to vanish at the horizon.}. 
Thus the effect of the curved background on the Higgs potential is to 
replace $T$ by $T_{\ef}$ and add an additional term proportional to the 
curvature, i.e. the third term in \staticpot. The large temperature limit 
can be alternatively interpreted as the large radius limit for the boundary 
3-geometries. Notice that $T_{\ef} \rightarrow T$ as $T$ tends to infinity.
For completeness we give the expansion of the potential 
for values of the scalar mass $M << T_{\ef}$
\eqn\hightemp{ V  = {\pi^2 T_{\ef}^4 V_3 N \over 4} \sum_{k=1}^{\infty} 
\left( {M\over \pi T_{\ef}}\right)^{\! k} \Bigl( a_k + 
{k\over 2 \pi^2 T_{\ef}^2} b_k \Bigr) \,,}
where the coefficients are given by
\eqn\coeffs{ a_k = (-)^k {k+3 \choose k}\;,\;\; b_1 = -4\;,\;\; 
b_2 = 10\;,\;\; b_{k>2} = 3\,(-)^k(k+1) \,.}

\ifig\spherical{Static D3-brane potential in the background of spherical 
AdS-black holes for $T_{min}$, $T_{crit}$ and $T>T_{crit}$.
We have also plotted the potential obtained from empty AdS in the 
global patch.}
{
%\vskip5cm 
%\hskip10cm
\epsfxsize=4truein\epsfysize=2.5truein
\epsfbox{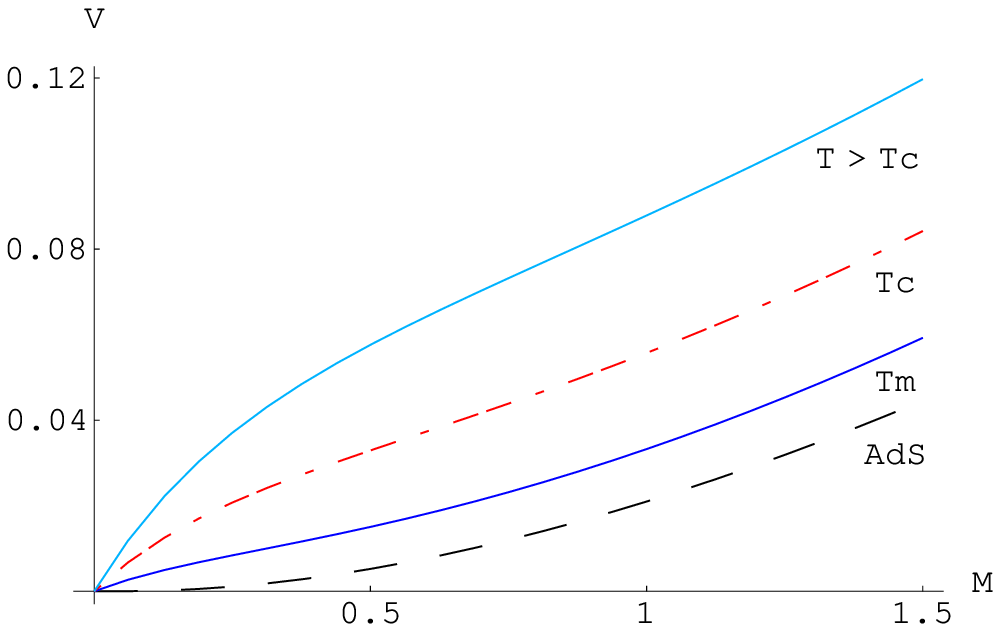}}

The definition \efftempi\ of $T_{\ef}$ allows us to treat also the 
case of empty AdS in the global patch by taking $T_{\ef} = 
{1\over \sqrt{2}\pi}$.
The D3-brane potential for $k=1$ at different temperatures 
as a function of the scalar mass $M$ is shown in \spherical. 
\ifig\hyperbolic{Static D3-brane potential in the background of 
hyperbolic AdS-black holes (on the vertical axes we have ploted $V/N V_3$).}
{%\vskip.1cm \hskip1.5cm
\epsfxsize=3.5truein\epsfysize=3truein
\epsfbox{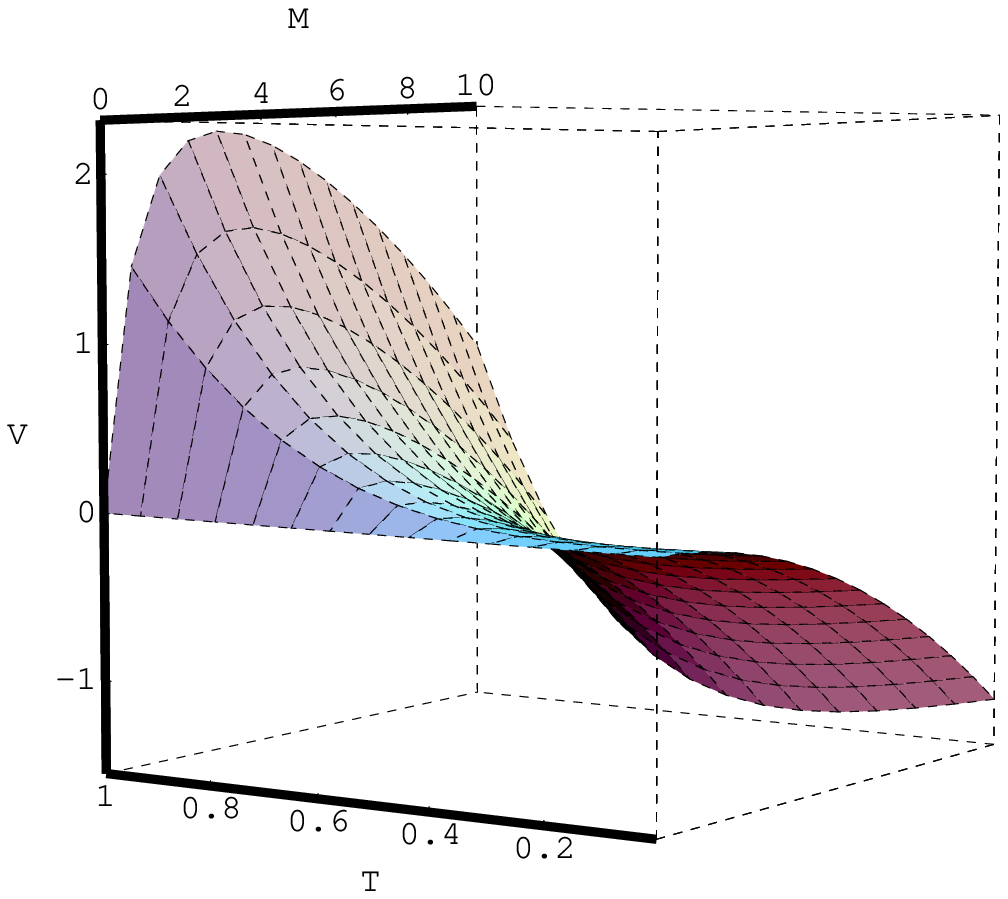}}

For $k=-1$, the conformal coupling to the curvature of the hyperboloid 
induces a negative mass term for the ${\cal N}=4$ scalars. We have already
seen in \largeu\ that this tree level coupling is reproduced in the D3-brane 
potential. Therefore we would expect the gauge theory to be unstable. 
An analysis of \staticpot\ shows however that the static potential
grows for small $M$ reaching a maximum at
\eqn\maxi{ M =  
\pi T_{\ef}\left({ 1+\left(2\pi^2 T_{\ef}^2 + 
\sqrt{4\pi^4 T_{\ef}^4-1}\right)^{2\over 3} \over
 \left(2\pi^2 T_{\ef}^2 + \sqrt{4\pi^4 T_{\ef}^4-1}\right)^{1\over 3}}
\right)^{1\over 2} -\pi T_{\ef}  \,.}
This formula is valid also for $T_{\ef} < {1\over \sqrt{2} \pi}$, where
${1\over \sqrt{2} \pi}$ is the effective temperature of empty AdS. 
The presence of a maximum means that in spite of the negative tree level mass the gauge theory 
is still driven back to the symmetric phase due to thermal effects as 
long as the scalar vacuum expectation value is not too large. Of course 
these statements apply to the limit of large $N$ and large 't Hooft coupling 
$\lambda$, where the supergravity approximation is valid. Notice that
the high of the potential maximum is proportional to $N$.
For finite $N$ and coupling the theory 
could still be unstable e.g. due to tunneling of D3-branes through the 
potential wall. It would also be interesting to see how much of this 
behavior is reproduced at weak coupling. For this one would have to do a 
one-loop calculation of the effective potential of ${\cal N}=4$ SYM on the 
hyperbolic background in the Higgs phase at finite temperature. A precise 
comparison of the gauge theory result with the results obtained here from 
supergravity seems problematic due to the well-known gauge dependence of the 
effective potential. The values of the effective potential at extrema 
are however gauge independent and can in principle be compared with the 
D3-brane potential at the maxima \maxi.

For $T_{\ef} \rightarrow 0$ the maximum of the potential tends to $M=0$. At
$T_{\ef}=0$ the potential is monotonically decreasing with 
$M$. 
In fact the D3-brane potential at zero temperature is precisely given by the 
tree level potential. This suggests 
that the field theory at zero temperature is unstable, a feature that,
according to the AdS/CFT conjecture, should be shared by the dual 
supergravity solution. 
In \gary\ a stability analysis of the extremal case has been presented. 
The authors found however that this geometry is stable against small 
perturbations. The resolution of this puzzle might go as follows.
A D3-brane probe on the extremal hyperbolic black hole
will be driven away from the horizon unless it is placed exactly at
$u_+$, the unstable maximum of the potential.  Since the length scale of
the supergravity solution is $\ell^4 \sim N$, it seems that the extremal
geometry is unstable against decreasing $\ell$ and therefore increasing
its curvature. Therefore one should possibly not only consider variations in 
the metric but also allow for changes of the four-form potential that 
determines the five-form flux on the $S^5$.

\newsec{Wilson Loops}

We will analyze now the behavior of Wilson loops and heavy 
quark potentials of ${\cal N}=4$ Yang-Mills in $S^3$ and $H^3$ using the 
AdS/CFT correspondence. At large $N$ and large 't-Hooft coupling, the 
expectation value of a Wilson loop $C$ is given in terms of the dual 
string theory by \refs{\jw, \reyw} 
\eqn\wl{
\langle W(C) \rangle \sim e^{-(S(C) - S_0)},\;\;\;\;\;\; 
}
where $S(C)$ is the Nambu-Goto action of the fundamental string whose 
world-sheet ends on the contour $C$ at the $u \rightarrow \infty$ boundary 
and minimizes the action. We will consider a contour describing 
the world-line of a $q \bar{q}$ pair of very heavy quarks. The 
action $S(C)$ is infinite due to the contribution of the world-sheet 
region close to the boundary. This divergence represents the self-energy 
of the non-dynamical quarks. In AdS language a single external quark is 
represented by a string extending straight along the radial coordinate 
into the AdS interior. It is therefore natural to regularize $S(C)$ by 
subtracting the action of two separated strings stretching down to $u=0$ in 
the case of empty AdS \refs{\jw, \reyw}, or to the horizon for black hole 
backgrounds \refs{\ya, \theisenwlt}. We have denoted this term by $S_0$ in \wl.

We will restrict ourselves to the simplified case where $q$ and $\bar q$ 
have equal charge under the ${\cal N}=4$ R-symmetry group, $SU(4)$. This 
implies that the minimal area configuration will lie at a single point 
in the $S^5$ dimensions. We take $q$ and ${\bar q}$ to be static and
using the symmetries of the sphere and hyperboloid we place them at 
$(\theta,\phi, \psi)=(0,0,0)$ and $(\theta,0,0)$ (for the sphere 
$\theta \in (0,\pi)$ while for the hyperboloid $\theta \in (0,\infty)$). 

The energy of the 
$q {\bar q}$ configuration is given by the vacuum expectation value of the Wilson loop
divided by $\beta$.
Following  \jw\ - \theisenwlt\ we find\foot{Recently deformations of AdS 
black holes with non-constant dilaton and the quark-antiquark potential 
in these backgrounds has been studied in \odintsov.}
\eqn\Vsh{
E =  {\lambda^{1\over 2} \over \pi z  } \left[ 
\int_1^{\infty} dy \left( \sqrt{ {y^4 + k z^2 y^2 - \mu z^4
\over y^4 + k z^2 y^2 - 1 -k z^2}} -1 \right)
-1 + z u_+ \right]. 
}
We have defined $z=1/u_0$ with $u_0$ being the minimal 
radial coordinate of the world-sheet ending on $C$. 
The mass of the black hole is related 
to the effective temperature defined in \efftempii\ by $\mu=
(\pi T_{\ef} )^4 -1/4$. As in the previous section, expression 
\Vsh\ describes both the cases of an $S^3$ and $H^3$ boundary by taking 
respectively 
$k=1$ or $k = -1$. The angular distance between the quarks is given 
in terms of $z$ by
\eqn\angle{
\theta = 2 z \sqrt{ 1 + k z^2 - \mu z^4} 
\int_1^{\infty}\!\!
{dy \over \sqrt{\bigl(y^4 + k z^2 y^2 - \mu z^4 \bigr)
\bigl( y^4 +k z^2 y^2 - 1 -k z^2 \bigr)}}.
}
The interaction potential between the pair of static quarks can be
read directly from \Vsh. If $E \leq 0$, $V_{q \bar{q}}(\theta)=E$. 
When $E>0$ the configuration with two separated world-sheets 
extending straight to the AdS interior is energetically favored. 
Then $V_{q {\bar q}}=0$ corresponding
to screening of the charges \refs{\ya, \theisenwlt}.

\subsec{Quark-antiquark potentials on $S^3$.}

The most interesting aspect of ${\cal N}=4$ Yang-Mills on $S^3$ is 
the existence of a phase transition at $T=T_{crit}$ \wsec.
In the supergravity dual its counterpart is  
the Hawking-Page phase transition between black hole and pure AdS 
solutions to the Einstein's equations \hp. The study of quark-antiquark 
potentials should provide an additional piece of information about the 
physics involved in it.

For temperatures below $T_{crit}$ the supergravity dual to ${\cal N}=4$ 
Yang-Mills on $S^3$ is empty AdS with the euclidean time coordinate $\tau$ 
periodically identified with period $\beta = 1/T$. 
The low temperature phase corresponds to set $\mu=0$,
$k=1$ in \Vsh\ and \angle, independently of $T$. Those expressions can be
integrated to give\foot{Quark-antiquark potentials in a Yang-Mills state 
associated to the uniform 
distribution of flat 3-branes on a disk have been studied recently in \ak. 
When the $q {\bar q}$ pair is placed on an axes orthogonal to the disk, 
the resulting expressions coincide formally with (3.4).
Since the Nambu-Goto action depends 
on products of metric elements, different metrics can produce formally
equivalent expressions for the Wilson loops.} 
\eqn\edzero{
\eqalign{
\theta = & {2 z \over \sqrt{(1+z^2)(2+z^2)}} \left( \Pi(\kappa'^2,\kappa)-K(\kappa) 
\right) , \cr  
E = & {\lambda^{1\over 2} \sqrt{2+z^2} \over  \pi z} \left( \kappa'^2 K(\kappa)-
E(k) \right),}
}
where $\kappa=1/\sqrt{2+z^2}$, $\kappa'=\sqrt{1-\kappa^2}$. The functions $K(\kappa)$, 
$E(\kappa)$ 
and $\Pi(\alpha^2,\kappa)$ denote the complete elliptic integrals of first, 
second and third kind \foot{We follow the notation in \elliptic.}.
For small $z$, which corresponds both to small inter-quark separations 
and to the large radius limit of $S^3$, we recover the flat limit result 
\refs{\jw, \reyw}
\eqn\Vflat{
V_{q {\bar q}}= { \lambda^{1\over 2} \over L \pi} \left( 
\Pi( {\textstyle{1 \over 2}},{\textstyle{1 \over \sqrt{2}}} ) - 
K({\textstyle{1 \over \sqrt{2}}}) \right)
\left( K ( {\textstyle{1 \over \sqrt{2}}} ) -
2 E( {\textstyle{1 \over \sqrt{2}}}) \right)=
-{4 \pi^2 \lambda^{1/2} \over \Gamma \left( {1 \over 4} \right)^4 L}.
}
Since we are considering 3-geometries of unit radius 
the inter-quark separation is just $L= \theta$.

\ifig\Vzero{(a) Potential for a pair of static heavy quarks on $S^3$
on the large $N$, large 't-Hooft coupling limit
(we have set $\lambda\!=\!1$). (b) Normalized ratio between the strong 
and weak coupling limits of the potential.}
{%\vskip.1cm \hskip1.5cm
\epsfxsize=7truein\epsfysize=2truein
\epsfbox{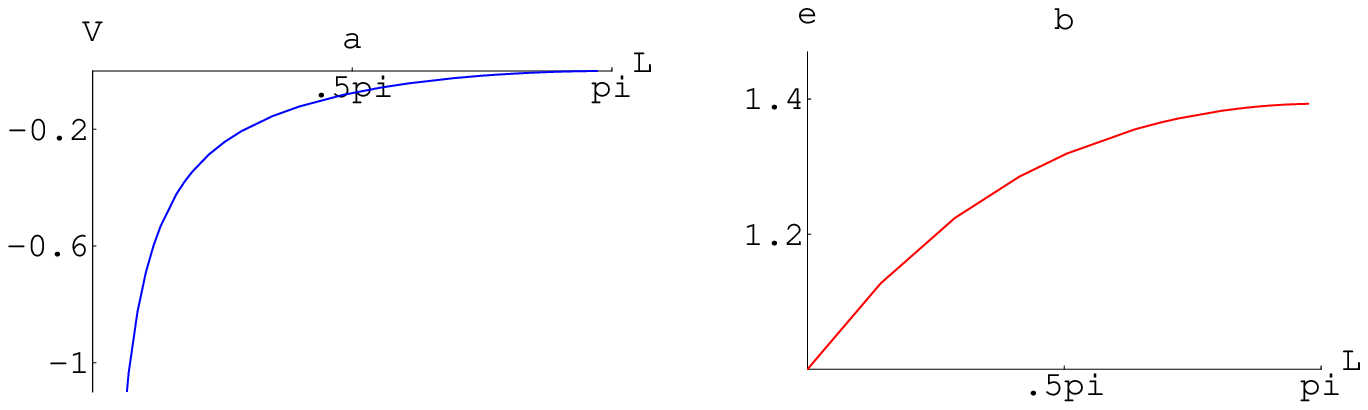}}

The fact that the entropy of empty AdS is zero indicates that the $N^2$ 
degrees of freedom of the dual gauge theory are somehow ``confined''. 
In \wsec\ it was argued that this is not dynamical confinement but just 
a kinematical effect: Gauss law on a compact space does not allow for net 
charges. In agreement with that, \Vzero\ (a) shows that the strong coupling 
quark-antiquark potential has a Coulomb-like behavior. Kinematical
confinement is also present at weak coupling 
since the ${\cal N}=4$ gauge theory 
is in a Coulomb phase at zero temperature. 

We will now compare the weak and strong coupling
limits of the potential more quantitatively. 
In the weak coupling and zero temperature limit, the potential between 
a pair of static charges on $S^3$ can be calculated explicitly. 
At small coupling we can neglect the non-linearity of Yang-Mills and
just consider an abelian ${\cal N}=4$ gauge theory. The potential between 
a pair of static charges with equal R-charge is due to the interchange of 
gauge bosons and a linear combination of the six ${\cal N}=4$ scalars, i.e. 
that with R-charge aligned with the quarks R-charge 
\eqn\Vweak{
V_{w}= - { g_{YM}^2 N \over 4 \pi^2 } \left( (\pi-\theta)\, \cot \theta 
+ {(\pi-\theta)\over  \sin \theta } \right).
}
The first term is associated to gauge boson interchange. It is 
obtained by solving 
the Laplace equation on $S^3$ with opposite sign delta sources at the 
positions of the quarks, and deriving from that the energy of the 
configuration. The second term in \Vweak\ is 
due to the scalar interchange. It is obtained by solving the
Laplace equation with sources and a mass term since the ${\cal N}=4$ 
scalars couple to the curvature of the 3-sphere. Indeed we have seen in 
the previous section that the scalars are conformally coupled, i.e. 
$m^2=1$, as expected \wfirst.  

The comparison between the weak and strong coupling limits of the 
potential at $T=0$ shows important renormalization effects. As in the 
flat case \jw, the dependence of the potential on the 't-Hooft coupling 
changes from weak to strong coupling. On $S^3$, in addition, also its 
functional dependence on the inter-quark separation changes. The radius of 
the sphere introduce a new scale and even in the conformally coupled 
case nothing constrains the dependence of the potential on the angular 
separation of the quarks. This does not contradict the fact that 
empty AdS is an exact solution to the string equations of motion \bg\ 
since \Vsh\ and \angle\ can receive $\alpha'$ corrections due 
to quantum fluctuations of the string ending on the quarks world-line 
\theisen. In order to compare the different $\theta$ dependence of 
\edzero\ and \Vweak, we have plotted in \Vzero\ (b) the quotient 
between the strong and weak coupling potentials normalized  
such that it tends to 1 at $\theta \rightarrow 0$, i.e. 
$e=e_0 V_{st}/V_{w}$ with 
\eqn\nor{
e_0=  \lim_{\theta \rightarrow 0} {V_{w} (\theta)
\over V_{st}(\theta)}= { \Gamma(1/4)^4 \lambda^{1/2}
\over 2 (2 \pi)^3}.
}
The potential at strong coupling is proportional to $\lambda^{1/2}$
instead of $\lambda$, which is the weak coupling result. It is interesting
to notice that this reduction in the expected intensity of the 
potential at strong coupling decreases slightly as the angular 
separation of the quarks grows. 

At $T \geq T_{crit}$ the field theory undergoes a phase transition,
detected because its supergravity dual description changes from
empty AdS to an AdS Schwarzschild black hole background. The black hole 
solutions 
have an entropy $S \sim N^2$. This implies that there are $\sim N^2$ 
degrees of freedom contributing to the entropy of the dual gauge theory 
and suggests that the gauge theory is in a deconfined phase \refs{\wsec, 
\wfirst}. Since low temperature confinement is just the statement of Gauss 
law on the sphere, we could think that what triggers the phase transition 
is screening. Let us analyze the situation at weak coupling. 
The $q \bar q$ potential at weak coupling is proportional to $\exp(-m_{el} 
L) \over L$. To lowest order we can estimate the electric screening mass 
to be $m_{el}^2 \propto\lambda T^2$ \kapusta. If the screening length 
$1\over m_{el}$ is much smaller than the size of the compact manifold the 
gauge theory will qualitatively behave as in flat space. Charges will be 
screened and we expect the entropy to scale like $N^2$. If however the 
screening length is bigger or of the order of the compact space size, the 
quark charges are not screened over maximal distances on the compact space. 
In this case physical states have to be color neutral and no quasi-free 
charges are possible. Thus we expect the entropy to scale like $N^0$. 
Therefore even at weak coupling there will be a phase transition in
the large $N$ gauge theory at finite temperature on a compact space.
If there is not other dominant mechanism, screening phenomena will
induce a phase transition at $T_c \propto {1\over \sqrt{\lambda} \pi}$ for 
the case of a three sphere of unit radius where the maximal distance is 
$\pi$. Although at weak coupling this crude estimate gives a very high
critical temperature, strong coupling effects could importantly lower it.

\ifig\figu{Energy of the single string configuration (a) and 
quark-antiquark separation (b) as functions of the minimal radial 
distance that the string reaches in the black hole background. 
(H denotes the black hole horizon).}
{%\vskip.1cm \hskip1.5cm
\epsfxsize=7truein\epsfysize=2truein
\epsfbox{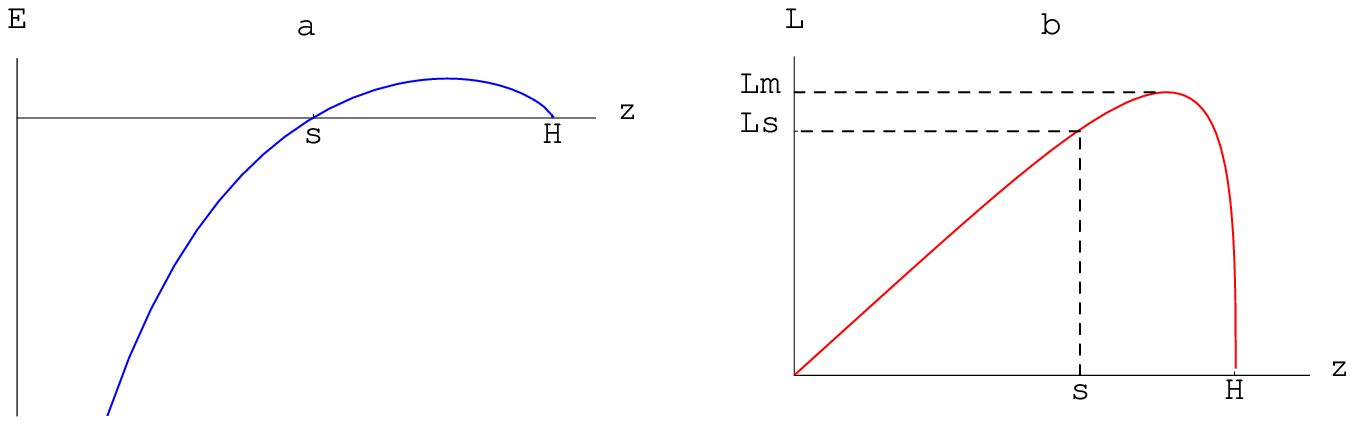}}

We derive again $V_{q {\bar q}}$ from \Vsh\ and \angle. The high
temperature phase of ${\cal N}=4$ on $S^3$ corresponds to $\mu
\geq 2$. The results
parallel those obtained in the flat case \ya\ \theisenwlt. The single 
string configuration with boundary on the quarks positions exists only 
for inter-quark distances smaller than a certain threshold, $L_m (T)$.
For each distance $L<L_m(T)$ there are two string configurations 
that extremize the Nambu-Goto action. One of them has bigger action 
than that of two separated strings extending straight to the black hole
horizon and is therefore energetically disfavored. For separations 
$L_s(T)<L \leq L_m(T)$, 
the action of the second single string configuration is also bigger than 
that of two separated strings (see \figu). Therefore $V_{q {\bar q}}=0$ 
for quark separations bigger than $L_s(T)$: the high temperature phase 
exhibits total screening.

\ifig\Vt{Quark-antiquark potential on $S^3$ for $T>T_{crit}$, $T_{crit}$
and $T_{min}$ ($\lambda=1$). In all cases the screening length is much smaller 
than $\pi$.}
{%\vskip.1cm \hskip1.5cm
\epsfxsize=4truein\epsfysize=2.5truein
\epsfbox{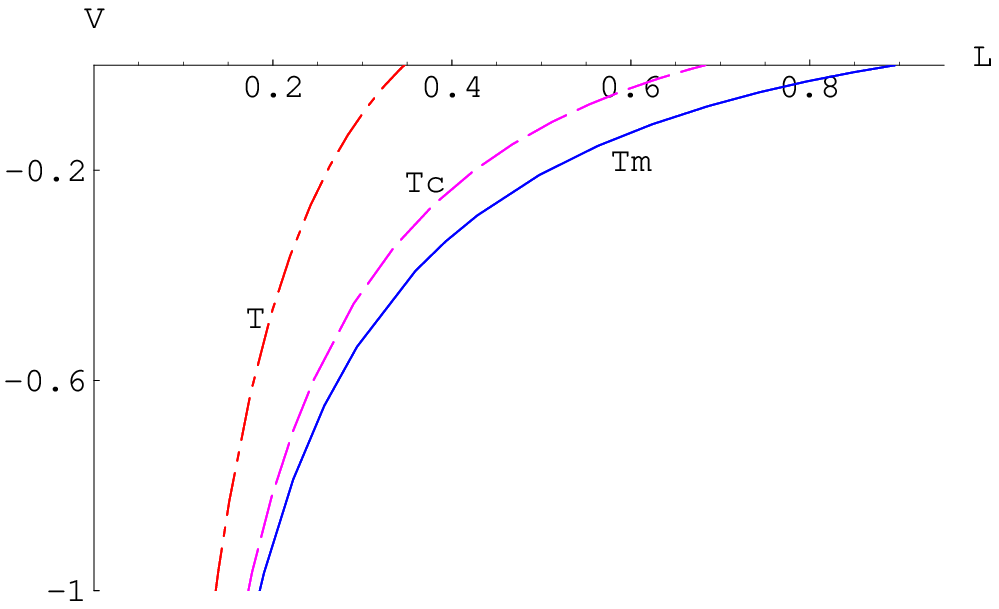}}
We observe in  \Vt\ that the screening length is always much smaller 
than $ \pi$. AdS Schwarzschild 
black hole solutions exist for $T\geq T_{min}$ though they are only 
stable for temperatures bigger than $T_{crit}>T_{min}$. Black hole
solutions with $T_{crit}>T \geq T_{min}$ could correspond to 
meta-stable states in the dual gauge theory. For completeness, we have
also plotted in \Vt\ the potential induced by the black hole
solution at $T_{min}$. The screening length is still very small.
Since the supergravity solutions we are treating have constant
dilaton, the results obtained for the screening length between
electric charges apply also to magnetic and dyonic charges.
According to this result the phase transition is not 
triggered by screening of the electric or magnetic charges. If the degrees 
of freedom at
strong coupling were electrically charged their charges would still
be completely screened even at (and below) the critical temperature.
Our results seem therefore to indicate that the relevant $N^2$
degrees of freedom at the phase transition do not carry electric charges.  
The other possible interpretation is that kinematical confinement of the
charges is not the mechanism which triggers the phase transition.

\subsec{Quark-antiquark potentials on $H^3$.}

\ifig\Vh{Quark-antiquark potentials on $H^3$ at $T=0$, $T_{AdS}$
and $T > T_{AdS}$.}
{%\vskip.1cm \hskip1.5cm
\epsfxsize=4truein\epsfysize=2.5truein
\epsfbox{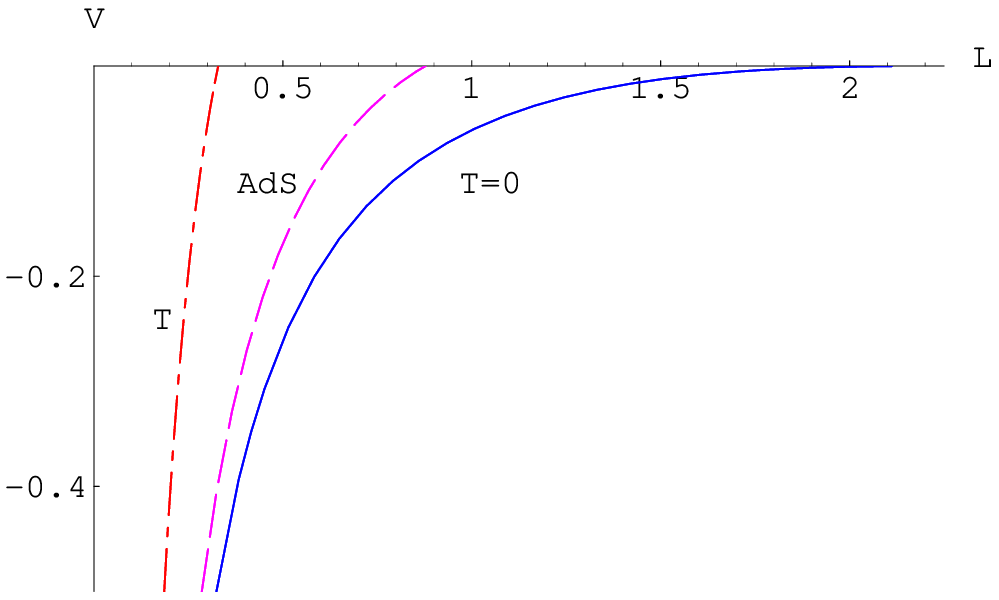}}
We turn now to analyze quark-antiquark potentials of ${\cal N}=4$
Yang-Mills on the 3-hyperboloid. In the large $N$ and large 't-Hooft 
coupling limit the inter-quark potential is derived from \Vsh\ and 
\angle\ by setting $k\!=\!-1$ and $\mu \! \geq \!-1/4$; the results 
are plotted in \Vh. 
As in the flat case, for temperatures bigger than zero there is
a maximal quark separation after which the charges are completely
screened. The surprising result is that even at $T\!=\!0$ the potential 
has a finite range! 

For $T>0$ the dependence of $V_{q {\bar q}}$ and the inter-quark
separation on the minimal radial distance of the single string
configuration is as shown in \figu. In particular this 
is true for $T=1/ 2 \pi $, when the background geometry is just
empty AdS with a hyperbolic boundary. The situation at $T=0$ is 
however different. The action of the single string 
configuration is always smaller than that of two separated strings. 
For each inter-quark separation there is only one single string 
configuration. As the tip of the string gets closer to the black
hole horizon, the inter-quark separation tends to the value
\eqn\ls{
L_s= \sqrt{2}  \int_0^{\infty} \! {dx \over (x+1)\sqrt{x (x+2)}}=
{\pi \over \sqrt{2}}.
}
This relation is obtained by changing variables $y^2\!-\!1\!=\!(1\!-\!{z^2\over 2})\, 
x$ in \angle\ with $z\! <\! 1$ and $\mu\!=\!0$, and then taking the limit 
$z \rightarrow 1$ \foot{As for the low temperature 
phase on $S^3$, \Vsh\ and \angle\ with $k\!=\!-1$ and $\mu\!=\!-1/4$ 
can be integrated in terms of elliptic functions. The results coincide 
formally with those of \ak\ when the $q\bar q$ pair lies in the plane of 
the disk (see footnote 6).}. 
For 
separations bigger than $L_s$ the only possible configuration is that of 
two separated strings, i.e. the charges are screened. 
Of course since we are on a hyperbolic space, the Coulomb potential
will also show exponential falloff at zero temperature in the weak coupling limit,
$V \sim \coth \theta$. The result at strong coupling deviates however 
from this weak coupling behavior since we find a maximum interaction distance.
This indicates an additional screening of the
charges at strong coupling\foot{Screening at $T=0$ has been also encountered in 
\ak. The geometry describing that case 
coincides with the extremal limit of a rotating D3-brane solution
\rot\ \rots. It is interesting to notice that the hyperbolic black 
hole geometry at $T=0$ is also extremal.}. 
The black hole background at $T=0$ has zero energy but non-zero entropy 
$S \sim N^2$ \rcm\ \roberto. The presence of these $N^2$ states
contributing to the entropy but not to the energy could be related with 
the presence of total screening.

\vskip1cm

\centerline{\bf Acknowledgments}
\vskip2mm
    
The work of K.L. was supported by an FWF project under number P13125-TPH.
The work of E.L. was supported by an FWF project under number P13126-TPH.

\listrefs
\end